\documentclass[twocolumn,preprintnumbers,amsmath,amssymb, bibnotes,nofootinbib]{revtex4}
\usepackage{amsthm}
\usepackage{amsmath}
\usepackage{amssymb}
\usepackage{graphicx}
\usepackage{url}
\usepackage{dsfont}
\usepackage{float}
\usepackage{bm}
\usepackage{ifthen}
\usepackage[usenames,dvipsnames]{color}
\usepackage{mathrsfs}
\usepackage[colorlinks=true,citecolor=blue,urlcolor=black]{hyperref}
\usepackage{float}
\usepackage{afterpage}
\usepackage{subfigure}
\usepackage{times}

\newcommand{\eg}{{\it{e.g.}}}
\newcommand{\ie}{{\it{i.e.}}}

\newcommand{\be}{\begin{equation}}
\newcommand{\ee}{\end{equation}}


\newcommand{\sket}[1]{{\ensuremath{\lvert#1\rangle}}}
\newcommand{\lket}[1]{{\ensuremath{\left\lvert#1\right\rangle}}}
\newcommand{\ket}[1]{\if@display\lket{#1}\else\sket{#1}\fi}
\newcommand{\sbra}[1]{{\ensuremath{\langle#1\rvert}}}
\newcommand{\lbra}[1]{{\ensuremath{\left\langle#1\right\rvert}}}
\newcommand{\bra}[1]{\if@display\lbra{#1}\else\sbra{#1}\fi}
\newcommand{\sbraket}[2]{{\ensuremath{\langle#1\rvert#2\rangle}}}
\newcommand{\lbraket}[2]{{\ensuremath{\left\langle#1\!\left\rvert\vphantom{#1}#2\right.\!\right\rangle}}}
\newcommand{\braket}[2]{\if@display\lbraket{#1}{#2}\else\sbraket{#1}{#2}\fi}

\newcommand{\sketbra}[2]{{\ensuremath{\lvert #1\rangle\!\langle #2\rvert}}}
\newcommand{\lketbra}[2]{{\ensuremath{\left\lvert #1\right\rangle\!\!\left\langle #2\right\rvert}}}
\newcommand{\ketbra}[2]{\if@display\lketbra{#1}{#2}\else\sketbra{#1}{#2}\fi}

\theoremstyle{plain}

\theoremstyle{definition}

\begin{document}
\title{Protocol choice and parameter optimization in decoy-state measurement-device-independent quantum key distribution}

\author{Feihu Xu$^{1}$}
 \email{feihu.xu@utoronto.ca}
\author{He Xu$^{1}$}
\author{Hoi-Kwong Lo$^{1}$}

\affiliation{%
$^{1}$ Centre for Quantum Information and Quantum Control, Department of Physics and Department of Electrical \& Computer Engineering, University of Toronto, Toronto,  Ontario, M5S 3G4, Canada }

\date{\today}
\begin{abstract}
Measurement-device-independent quantum key distribution (MDI-QKD) has been demonstrated in both laboratories and field-tests using attenuated lasers combined with the decoy-state technique. Although researchers have studied various decoy-state MDI-QKD protocols with two or three decoy states, a clear comparison between these protocols is still missing. This invokes the question of how many types of decoy states are needed for practical MDI-QKD. Moreover, the system parameters to implement decoy-state MDI-QKD are only partially optimized in all previous works, which casts doubt on the actual performance of former demonstrations. Here, we present analytical and numerical decoy-state methods with one, two and three decoy states. We provide a clear comparison among these methods and find that \emph{two} decoy states already enable a near optimal estimation and more decoy states cannot improve the key rate much in either asymptotic or finite-data settings. Furthermore, we perform a full optimization of system parameters and show that full optimization can significantly improve the key rate in the finite-data setting. By simulating a real experiment, we find that full optimization can increase the key rate by more than one order of magnitude compared to non-optimization. A local search method to optimize efficiently the system parameters is proposed. This method can be four orders of magnitude faster than a trivial exhaustive search to achieve a similar optimal key rate. We expect that this local search method could be valuable for general fields in physics.
\end{abstract}
\maketitle

\section{Introduction}~\label{Sec:introcution}
Quantum cryptography or quantum key distribution (QKD) can provide information-theoretic security based on the laws of quantum physics~\cite{bennett1984quantum,ekert1991quantum}. During the past decade, commercial QKD products have appeared on the market; various field-test QKD networks have already been built around the world~\cite{peev2009secoqc,sasaki2011field}. However, owing to the imperfections in real-life implementations of QKD, a gap between its theory and practice remains unfilled. In particular, an eavesdropper (Eve) may exploit these imperfections and launch attacks not covered by the original security proofs of QKD. Indeed, the recent demonstrations of various attacks~\cite{Yi:timeshift:2008, Xu:phaseremapping:2010, Lars:nature:2010, Gerhardt:2010, weier2011quantum,jain2011device} on top of practical QKD systems highlight that this gap is a major problem for the real-life security of QKD.

Fortunately, measurement-device-independent quantum key distribution (MDI-QKD)~\cite{Lo:MDIQKD} removes all detector side-channel attacks, the most important security loophole in conventional QKD implementations~\cite{Yi:timeshift:2008,Xu:phaseremapping:2010, Lars:nature:2010, Gerhardt:2010, weier2011quantum,jain2011device}. The key idea of MDI-QKD is that both legitimate users (Alice and Bob) are senders. They transmit signals to an \emph{untrusted} third party, Charles (or Eve), who is supposed to perform a Bell state measurement. Such a measurement provides post-selected entanglement that can be verified by Alice and Bob. By using a decoy-state protocol~\cite{Hwang:2003, Lo:2005, Wang:2005,ma2005practical,zhao2006experimental,rosenberg2007long,peng2007experimental,dixon2008gigahertz,wei2013decoy,lucamarini2013efficient,lim2013concise}, Alice and Bob can use imperfect single-photon sources such as attenuated lasers and still estimate the contributions from single-photon signals. Unlike security patches~\cite{yuan2011resilience} and (full) device-independent QKD~\cite{mayers2004self, acin2007device,braunstein2012side}, MDI-QKD can remove all
detector side channels and is also practical for current technology. Hence, it has attracted a lot of scientific attention from the research community on both theoretical~\cite{ma2012statistical,wang2013three,yu2013generalized,sun2013practical,Feihu:practical,marcos:finite:2013,xu2013long} and experimental~\cite{Tittel:2012:MDI:exp, Liu:2012:MDI:exp, da2012proof, zhiyuan:experiment:2013} studies.

Various decoy-state methods have been proposed for MDI-QKD. Ref.~\cite{ma2012statistical} proposed a numerical method using two and three decoy states. Refs.~\cite{wang2013three,yu2013generalized,sun2013practical,Feihu:practical,marcos:finite:2013} discussed different analytical approaches based on two decoy states. Experimentally, Refs.~\cite{Tittel:2012:MDI:exp,da2012proof,zhiyuan:experiment:2013} implemented the two decoy-state protocol, while Ref.~\cite{Liu:2012:MDI:exp} chose the three decoy-state protocol. Consequently, the first open question is: how many types of decoy states are essentially needed for MDI-QKD in practice?

On the other hand, to implement MDI-QKD, one has to know the parameters to optimize the system performance. However, some previous theoretical studies~\cite{ma2012statistical,wang2013three,yu2013generalized} and experimental implementations~\cite{Liu:2012:MDI:exp,da2012proof} simply choose empirical parameters \emph{without} optimization. Hence, an important question is: how can one optimize the parameters used in MDI-QKD? This question is non-trivial, given the large number of parameters involved. Another question is: how much will a careful parameter optimization improve performance?

It is well known that an efficient version of decoy-state BB84 with biased basis choice rather than the standard one can significantly improve the key rate~\cite{wei2013decoy, lucamarini2013efficient,lim2013concise}. In biased basis choice, the basis-sift factor can be 1 instead of 1/2 for the standard one, hence the maximal improvement is about 100\%~\cite{lo2005efficient}. In MDI-QKD, the $\rm X$ basis cannot be used to generate secure keys due to its large error rate~\cite{Lo:MDIQKD}, thus the basis-sift factor can be improved from 1/4 (standard one) to 1 (biased basis choice). An efficient protocol with biased basis choice has a \emph{larger} improvement up to 300\%.

The parameters to implement this efficient protocol are chosen via optimization. Previously, this optimization has been studied and implemented on decoy-state BB84~\cite{ma2005practical,zhao2006experimental,rosenberg2007long,peng2007experimental,dixon2008gigahertz, wei2013decoy, lucamarini2013efficient,lim2013concise} as well as decoy-state MDI-QKD~\cite{sun2013practical,Feihu:practical,marcos:finite:2013,xu2013long,Tittel:2012:MDI:exp,zhiyuan:experiment:2013}. Nonetheless, except for Refs.~\cite{wei2013decoy,lim2013concise}, it is only partial. That is, the basis choice is independent of the intensity choice, which we will call \emph{simplified choice} or partial optimization in this paper. This choice is simple for implementation as the sender's two modulators on the intensity and bit information can be completely independent. However, from the theoretical point of view, the simplified choice cannot result in the optimal key rate due to the finite-data effect. The \emph{optimal choice} should select the majority of signal state in the $\rm Z$-basis for key generation, while the majority of decoy states in the $\rm X$-basis for a good estimation on the phase error rate, \ie, the basis choice should depend on the intensity choice.

In the asymptotic case with infinite data-set, the optimal choice and the simplified choice are the same. In BB84, the optimal choice cannot improve the key rate much, because: a) it is relatively easy to generate a large amount of detection counts (approaching asymptotic case) using a high-speed system~\cite{lucamarini2013efficient}; b) the receiver (Bob) cannot implement the optimal choice as he cannot distinguish the signal state from decoy sates. In MDI-QKD, we note however that the optimal choice can significantly increase the key rate, because: a) the detection counts are relatively low\footnote{This is because MDI-QKD requires two-fold coincidence events whereas decoy-state BB84 needs only single detection; Given that standard InGaAs single photon detectors in telecom wavelength have a rather low efficiency of say 15\%, for a fixed duration of experiment, the data size generated by MDI-QKD is substantially lower than that generated in decoy-state BB84. See the simulation and experimental results~\cite{Feihu:practical,marcos:finite:2013, Tittel:2012:MDI:exp, Liu:2012:MDI:exp, da2012proof, zhiyuan:experiment:2013} for more details.}, \ie, they are away from the asymptotic case; b) both Alice and Bob are the sender and can explicitly know the intensity choice. Indeed, by simulating a real experiment, we find that optimal choice can improve simplified choice about 200\% in a reasonable data-set (see Table~\ref{Tab:key:differentchoices}).

Implementing optimal choice requires a \emph{full} parameter optimization with the numerical search over many dimensions including the intensity choice of signal state and decoy states and the probability choice of intensities and bases. With a trivial exhaustive search, such an optimization problem is believed to be a challenge in terms of computational complexity (see Table~\ref{Tab:localsearch} as well as the case in decoy-state BB84~\cite{lim2013concise}). This might be one of the major reasons that a full parameter optimization is neglected in all previous works on decoy-state MDI-QKD~\cite{ma2012statistical,wang2013three,sun2013practical,yu2013generalized,Feihu:practical,marcos:finite:2013,xu2013long,Tittel:2012:MDI:exp, Liu:2012:MDI:exp, da2012proof, zhiyuan:experiment:2013}. Hence, another open question is: how one can perform a full parameter optimization in MDI-QKD?

In this paper, we provide solutions to the above open questions. We present analytical and numerical decoy-state methods with one, two and three decoy states. By clearly comparing these methods, we find that two decoy states combined with a full parameter optimization is already close to the optimal estimation and more decoy states cannot improve the key rate much in both asymptotic and finite-data cases. Moreover, we introduce a local search algorithm (LSA)~\cite{boyd2004convex}, a well-known algorithm in the field of computer science, to QKD for a full parameter optimization. This algorithm requires very low computational power and can be four orders of magnitude faster than a trivial exhaustive search to achieve a similar optimal key rate. It can also be applied to various decoy-state QKD protocols including MDI-QKD and BB84. Furthermore, we show that a full parameter optimization in MDI-QKD can improve the secure key rate more than one order of magnitude over non-optimization and it can still increase the key rate around 200\% over simplified choice in finite-data settings. Finally, we, for the first time, propose and experimentally implement a novel decoy-state method with only one decoy state. This method is simple to implement, but gives a slightly lower key rate. The protocol and the implementation results are respectively presented in Appendix~\ref{Sec:analytical} and~\ref{App:1decoy}.

The rest of this paper is organized as follows. We introduce the theory of decoy-state MDI-QKD in Sec.~\ref{Sec:decoyMDIQKD}. In Sec.~\ref{Sec:optimization}, we present our method to perform a full parameter optimization. We show the simulation results about the key rate comparison among full-optimization, partial-optimization and non-optimization in Sec.~\ref{Sec:numerical1}.  In Sec.~\ref{Sec:numerical2}, we present the simulation results for different number of decoy states. Finally, we conclude this paper in Sec.~\ref{Sec:conclusion}.

\section{Decoy-state MDI-QKD}~\label{Sec:decoyMDIQKD}
The secure key rate of MDI-QKD in the asymptotic case is given by~\cite{Lo:MDIQKD}
\begin{equation} \label{Eqn:Key:formula}
    R\geq P_{11}^{\rm Z}Y_{11}^{\rm Z}[1-H_{2}(e^{\rm X}_{11})]-Q^{\rm Z}_{\mu\mu}f_{e}(E^{\rm Z}_{\mu\mu})H_{2}(E^{\rm Z}_{\mu\mu}),
\end{equation}
where $Y^{\rm Z}_{11}$ and $e^{\rm X}_{11}$ are, respectively, the yield (the conditional probability that Charles declares a successful event) in the rectilinear ($\rm Z$) basis and the error rate in the diagonal ($\rm X$) basis, given that both Alice and Bob send single-photon states; $P_{11}^{\rm Z}$ denotes the probability that Alice and Bob send single-photon states in the $\rm Z$ basis; $H_{2}$ is the binary entropy function given by $H_2(x)$=$-x\log_2(x)-(1-x)\log_2(1-x)$; $Q^{\rm Z}_{\mu\mu}$ and $E^{\rm Z}_{\mu\mu}$ denote, respectively, the gain and QBER in the $\rm Z$ basis; $\mu$ is the intensity of the signal state and its optimal value in the asymptotic case is shown in Appendix~\ref{App:optimalmu}; $f_{e}\geq 1$ is the error correction inefficiency function. Here we use the $Z$ basis for key generation and the $\rm X$ basis for testing only~\cite{lo2005efficient}. In practice, $Q^{\rm Z}_{\mu\mu}$ and $E^{Z}_{\mu\mu}$ are directly measured in the experiment, while $Y^{\rm Z}_{11}$ and $e^{\rm X}_{11}$ can be estimated using the decoy-state methods~\cite{ma2012statistical,wang2013three,sun2013practical,Feihu:practical,marcos:finite:2013,yu2013generalized}.

In a MDI-QKD implementation with coherent states (attenuated lasers), by performing the measurements for different intensity settings, we can obtain~\cite{Lo:MDIQKD, Feihu:practical}
\begin{equation} \label{Eqn:general:decoy}
\begin{aligned}
& Q_{q_aq_b}^{\lambda}=\sum_{n,m=0}e^{-(q_a+q_b)}\frac{q^n_a}{n!}\frac{q^m_b}{m!}Y_{nm}^{\lambda},  \\
& Q_{q_aq_b}^{\lambda}E_{q_aq_b}^{\lambda}=\sum_{n,m=0}e^{-(q_a+q_b)}\frac{q^n_a}{n!}\frac{q^m_b}{m!}Y_{nm}^{\lambda}e_{nm}^{\lambda},
\end{aligned}
\end{equation}
where $\lambda\in\{\rm X,\rm Z\}$ denotes the basis choice, $q_a$ ($q_b$) denotes Alice's (Bob's) intensity setting, $Q_{q_aq_b}^{\lambda}$ ($E_{q_aq_b}^{\lambda}$) denotes the gain (QBER), and $Y_{nm}^{\lambda}$ ($e_{nm}^{\lambda}$) denotes the yield (error rate) given that Alice and Bob send respectively an $n$-photon and $m$-photon pulse. Here, the key idea of the finite decoy-state protocol is to estimate a lower bound for $Y^{\rm Z}_{11}$ and an upper bound for $e^{X}_{11}$ from the set of linear equations given by Eq.~(\ref{Eqn:general:decoy}). We denote these two bounds as $Y^{{\rm Z},L}_{11}$ and $e^{{\rm X},U}_{11}$ respectively.

In this work, we focus on the \emph{symmetric} case where the two channel transmissions from Alice to Charles and from Bob to Charles are equal. The analysis for asymmetric case can be equivalently conducted by following the techniques presented in~\cite{Feihu:practical}. In symmetric case, the optimal intensities for Alice and Bob are \emph{equal}~\footnote{In this symmetric case, one can prove that Alice's ($\mu_a$) and Bob's ($\mu_b$) optimal signal states in the asymptotic limit satisfy $\mu_a$=$\mu_b$ by using the model presented in the Appendix B and C1 of~\cite{Feihu:practical}; for practical settings, we have performed numerical simulations on all dimensions of parameters (\ie, $\mu_a$, $\nu_a$, $\omega_a$, $P_{\mu_{a}}$,... for Alice and $\mu_b$, $\nu_b$, $\omega_b$,$P_{\mu_{b}}$,... for Bob) and also find that the optimal intensities (and optimal probabilities) are equal.}. Hence, to simplify our discussion, we assume that equal intensities are used by Alice and Bob, \ie, $q_a$=$q_b$=$q$ with $q\in\{\mu,\nu_1,\nu_2,\omega, ...\}$, where $\mu$ denotes the signal state and $\{\nu_1,\nu_2,\omega, ...\}$ denote the decoy states.

This decoy-state estimation can be completed either numerically via linear programming~\cite{ma2012statistical,marcos:finite:2013} or analytically via gaussian elimination~\cite{wang2013three,sun2013practical,Feihu:practical,marcos:finite:2013,yu2013generalized}. The details of our numerical and analytical methods are respectively presented in Appendix~\ref{Sec:numerical} and~\ref{Sec:analytical}.

\section{Full parameter optimization for MDI-QKD} \label{Sec:optimization}

\begin{table}
\begin{tabular}{|c|c|c|c|}
  \hline
  Method & Iterations & Time & Key rate  \\
  \hline
  Exhaustive search & $10^{7}$ & 550 hours &  $6.84\times10^{-5}$ \\
  \hline
  Local search     & 33       & 1 min  & $6.83\times10^{-5}$ \\
  \hline
\end{tabular}
\caption{Comparison of local search local search algorithm (LSA) and exhaustive search. The simulation is conducted on MDI-QKD with two decoy-state numerical approach (Appendix~\ref{Sec:numerical}) using a standard desktop computer. A full optimization on eight dimensions including intensity and probability choices is performed. To reduce the computational complexity of exhaustive search, the intensity of $\omega$ is fixed at a near optimal value $\omega$=0.0005 (see Appendix~\ref{App:omega} for the general discussion about the effect of $\omega$). Exhaustive search applies 10 points on seven other dimensions and thus it requires $10^{7}$ iterations for optimization. LSA uses the coordinate descent and backtrack search algorithm~\cite{boyd2004convex}. LSA can not only maintain the accuracy of parameter optimization, but can also significantly reduce the computational complexity. The time needed for LSA is four orders of magnitude shorter than an exhaustive search.}~\label{Tab:localsearch}
\end{table}

In practical QKD applications, for better performance in terms of key rate and distance, it is advantageous to make a serious attempt to optimise operating parameters (\eg, signal/decoy state intensities, basis probabilities and signal/decoy state probabilities). As discussed in Sec.~\ref{Sec:introcution}, the simplified choice (or partial optimization) refers to the basis choice \emph{independent} of the intensity choice. It is commonly used in all previous works on MDI-QKD~\cite{ma2012statistical,wang2013three,yu2013generalized,sun2013practical,Feihu:practical,xu2013long,marcos:finite:2013,Tittel:2012:MDI:exp, Liu:2012:MDI:exp, da2012proof, zhiyuan:experiment:2013}. If $\rm Z$ is used as the majority basis for key generation, the simplified choice will modulate most of signals (over 90\%) on $\rm Z$ for all signal and decoy states. Nonetheless, the \emph{key} parameter in a decoy-state estimation is the bit error rate in $\rm X$, \ie, $e^{\rm X}_{11}$, which requires a large amount of detection counts for the decoy states in $\rm X$. The simplified choice, in contrast, results in a small number of such detection counts and thus increases the estimation error of $e^{\rm X}_{11}$ due to large statistical fluctuations. Therefore, the optimal choice (or full optimization) refers to the basis choice dependent on the intensity choice.

To perform this optimal choice in BB84 and MDI-QKD in the case of two decoy states, we are required to optimize two sets of parameters: intensities of signal and decoy states $\mu, \nu, \omega$, and the probabilities to choose different intensities and bases $P_{\mu}, P_{\nu}, P_{{\rm Z}|\mu}, P_{{\rm Z}|\nu}, P_{{\rm Z}|\omega}$, where $P_{\mu}$ denotes the probability to choose intensity $\mu$ and $P_{{\rm Z}|\mu}$ denotes the conditional probability to choose ${\rm Z}$ basis conditional on $\mu$. Essentially, it requires a search over eight dimensions~\footnote{In the case of Vacuum + weak decoy-state protocol~\cite{ma2005practical, wei2013decoy,wang2013three,sun2013practical}, the search can be reduced to six dimensions.}. Suppose that a trivial exhaustive search with 10 points on each dimension is conducted, it requires $10^{8}$ iterations to obtain the optimal parameters, which requires over 5000 hours on a standard desktop with 4-core CPUs~\footnote{Even though a monte carlo optimization is conducted on a high-performance computer such as the one with 16-core CPUs, it still requires a few days to complete such optimization~\cite{lim2013concise}.}. At first sight, it might appear to be a hard problem to perform full optimization. However, there is \emph{no} need to perform an exhaustive search.

Here, we introduce a local search algorithm (LSA)~\cite{boyd2004convex}, a well-known algorithm in the field of computer science, to QKD for this optimization problem. In particular, we adopt the coordinate descent and backtrack search algorithm~\cite{boyd2004convex} in our implementation. Coordinate descent can effectively transform a multi-dimensional optimization problem to a one-dimensional line search problem along the direction of one coordinate. This one-dimensional line search problem can be solved by backtrack search algorithm. As a consequence, the LSA enables one to perform a full optimization on all experimental parameters efficiently. We implement this LSA on MDI-QKD and show the comparison results to the trivial exhaustive search in Table~\ref{Tab:localsearch}. LSA can be four orders of magnitude faster than a trivial exhaustive search and also achieve a similar optimal key rate. Therefore, LSA cannot only reduce the computational complexity but also present a high accuracy. More details about this algorithm are shown in Appendix~\ref{App:Local}.

\begin{table}[hbt]
\centering
\begin{tabular}{c @{\hspace{0.5cm}} c @{\hspace{0.5cm}} c @{\hspace{0.5cm}} c @{\hspace{0.5cm}} c @{\hspace{0.5cm}} c} \hline
$\eta_{d}$ & $e_{d}$ & $Y_{0}$ & $f_{e}$ & $\epsilon$ & $N$ \\
\hline
14.5\% & 1.5\% & $6.02\times 10^{-6}$  & 1.16 & $10^{-7}$ & $10^{12}$ \\
\hline
\end{tabular}
\caption{(Color online) List of practical parameters for numerical simulations. These experimental parameters, including the detection efficiency $\eta_{d}$, the total misalignment error $e_{d}$ and the background rate $Y_{0}$, are from the 144 km QKD experiment reported in~\cite{Ursin:144QKD}. Since two SPDs are used in~\cite{Ursin:144QKD}, the background rate of each SPD here is roughly half of the value there. We assume that the four SPDs in MDI-QKD have identical $\eta_{d}$ and $Y_{0}$. $\epsilon$ is the security bound considered in our finite-data analysis. $N$ denotes the total number of signals (weak coherent pulses) sent by Alice and Bob.} \label{Tab:exp:parameters}
\end{table}

\section{Simulations on parameter optimization} \label{Sec:numerical1}
In all the simulations presented below, we use the experimental parameters, listed in Table~\ref{Tab:exp:parameters}, mostly from the long-distance QKD experiment reported in~\cite{Ursin:144QKD}.

\subsection{Key rate comparison between optimization and non-optimization}
\begin{figure}[!t]
\centering
\resizebox{8cm}{!}{\includegraphics{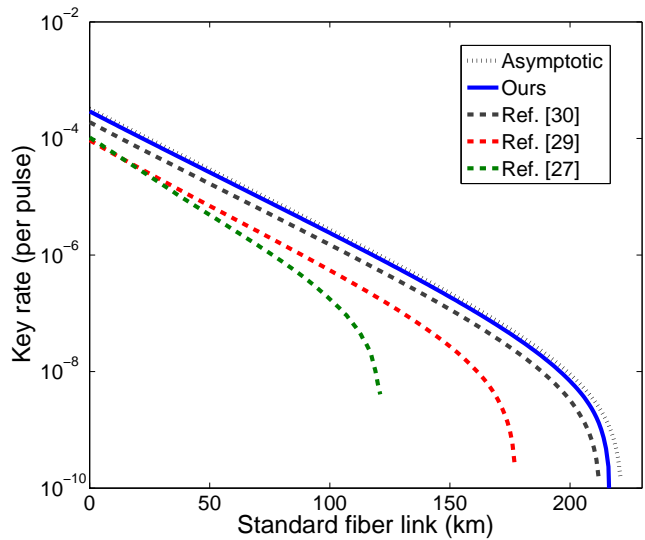}} \caption{(Color online) Key rate comparison with infinite data-set. The dotted black curve is the perfect key rate with infinite decoy states. The blue solid curve is our optimized key rate using the numerical approach with two decoy states, where the intensities are $\omega$=0.0005, $\nu$=0.01 and optimized $\mu$. For comparison purpose, we present the non-optimized and partially-optimized key rates using the methods and parameters of Refs~\cite{ma2012statistical,sun2013practical,yu2013generalized}: the black dashed curve is using~\cite{sun2013practical} with $\omega$=0, $\nu$=0.01 and optimized $\mu$; the red dashed curve is using~\cite{yu2013generalized} with $\omega$=0.01, $\nu$=0.1 and $\mu$=0.3; the green dashed curve is using~\cite{ma2012statistical} with $\omega$=0, $\nu$=0.1 and $\mu$=0.5. Notice that if the parameter optimization is also applied to Refs~\cite{ma2012statistical,yu2013generalized}, all the key rates are almost the same. In the asymptotic case, parameter optimization is simple, as only the intensities are required to be optimized and a smaller value of decoy-state intensity can in principle result in a better estimation. Parameter optimization can still increase the key rate and extend the secure distance.} \label{Fig:key:comp1}
\end{figure}

For previous works on decoy-state MDI-QKD, Refs.~\cite{ma2012statistical,yu2013generalized} used some typical parameters without optimization and Ref.~\cite{sun2013practical} performed a partial optimization only on intensity choice. Here, we first compare our optimized key rate to those using the parameters and methods presented in Refs.~\cite{ma2012statistical,yu2013generalized,sun2013practical}. Fig.~\ref{Fig:key:comp1} shows the comparison results in the asymptotic case. The dotted black curve is the perfect key rate with infinite decoy states. The blue solid curve is the key rate using our numerical method with two decoy states (see Appendix~\ref{Sec:numerical}), where we choose the near optimal intensities by maximizing the key rate~\footnote{Notice that in the asymptotic case, the key rate increases with the decrease of the intensity values of decoy states and the probability choice of intensities and basis is not required. To have a fair comparison to~\cite{sun2013practical}, we choose the same value of decoy state $\nu$ as $\nu=0.01$ and optimize $\mu$. These intensity values can already give a key rate close to the perfect key rate with infinite decoy state.}. The black, red and green dashed curves are respectively using the method and parameters of~\cite{sun2013practical},~\cite{yu2013generalized} and~\cite{ma2012statistical}. We can see that the key rates without parameter optimization in Refs~\cite{ma2012statistical, yu2013generalized} are much lower than ours and Ref.~\cite{sun2013practical}. Hence, parameter optimization not only increases the key rate but also extends the secure distance in the asymptotic case.

\begin{figure}[!t]
\centering
\resizebox{8cm}{!}{\includegraphics{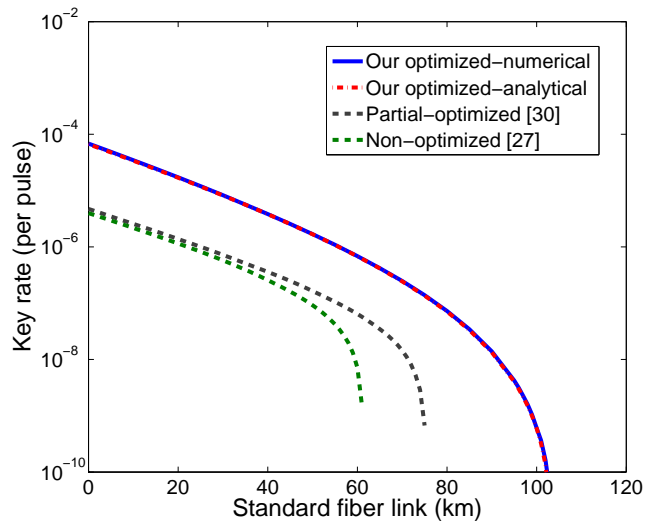}} \caption{(Color online) Practical key rate comparison (with statistical fluctuations). The optimal parameters and key rate in the distance of 50km (standard fiber) are shown in Table~\ref{Tab:optimalparameters}. All the key rates are simulated with $N$=$10^{12}$. The blue solid and red dashed-dotted curves (almost overlapped) are respectively our optimized key rates (after a full optimization) using the numerical (Appendix~\ref{Sec:numerical}) and analytical (Appendix~\ref{Sec:analytical}) methods with two decoy states. The black dashed curve is using the method of Ref.~\cite{sun2013practical}, where only partial parameters (\ie, the intensities) are optimized. The green dashed curve is using the method of Ref.~\cite{ma2012statistical}, where some typical parameters are assumed without optimization. Without full parameter optimization, the key rates in Refs~\cite{ma2012statistical,sun2013practical} are around one order of magnitude lower than ours across different distances. Our method can enable secure MDI-QKD over 25km longer than~\cite{ma2012statistical,sun2013practical}. These results highlight the importance of parameter optimization in practical decoy-state MDI-QKD.} \label{Fig:key:comp2}
\end{figure}

\begin{table}[hbt]
\centering
\begin{tabular}{c | @{\hspace{0.1cm}} c @{\hspace{0.2cm}} c @{\hspace{0.2cm}} c }
  \hline
   Parameters & Optimal & Ref.~\cite{ma2012statistical} & Ref.~\cite{sun2013practical}  \\
   \hline
  $\mu$      & 0.25 & 0.5 & 0.21 \\
  $\nu$      & 0.05 & 0.1 & 0.06 \\
  $\omega$   & $10^{-6}$ & 0 & 0  \\
  $P_{\mu}$  & 0.58 & 0.33 & 0.33 \\
  $P_{\nu}$  & 0.30 & 0.33 & 0.33 \\
  $P_{X|\mu}$  & 0.03 & 0.5 & 0.5 \\
  $P_{X|\nu}$  & 0.71 & 0.5 & 0.5 \\
  $P_{X|\omega}$  & 0.83 & 0.5 & 0.5 \\
  \hline
  $R$  & $1.68\times10^{-6}$ & $1.01\times10^{-7}$ & $1.64\times10^{-7}$   \\
  \hline
\end{tabular}
\caption{Comparison of parameters at 50km standard fiber. More general comparison results are shown in Fig.~\ref{Fig:key:comp2}. The 2nd column is the optimal parameters after a full parameter optimization. The 3rd and 4th columns are respectively the parameters from Refs.~\cite{ma2012statistical} and~\cite{sun2013practical}. We can see that full optimization can improve the key rate $R$ over one order of magnitude over the non-full-optimization of Refs.~\cite{ma2012statistical,sun2013practical}. This improvement mainly comes from optimizing the choices of intensities and probabilities. Notice that for the smallest decoy-state $\omega$, modulating the optimal value of around $10^{-6}$ is usually difficult in decoy-state QKD experiments~\cite{zhao2006experimental,rosenberg2007long,dixon2008gigahertz,lucamarini2013efficient}. However, we find that as long as the intensity of $\omega$ is below $1\times10^{-3}$, the key rate is very close to the optimum (see Appendix~\ref{App:omega} for details).} \label{Tab:optimalparameters}
\end{table}

Fig.~\ref{Fig:key:comp2} shows the practical key rates, \ie, with statistical fluctuations, in the case of data-size $N$=$10^{12}$. The optimal parameters and key rate for the distance of 50km (standard fiber) are shown in Table~\ref{Tab:optimalparameters}. Since Ref.~\cite{yu2013generalized} did not consider the finite-data effect, we do not show their key rate here. For a fair comparison, we use the method of standard error analysis~\cite{ma2012statistical} to analyze the statistical fluctuations. The key rates without full parameter optimization in Refs~\cite{ma2012statistical,sun2013practical} are around one order of magnitude lower than ours across different distances. Our method can enable secure MDI-QKD over 25km longer than~\cite{ma2012statistical,sun2013practical}. From Table~\ref{Tab:optimalparameters}, we can see that the improvement mainly comes from the optimization on the choices of intensities and probabilities. We have also performed such comparison at different data-sizes from $N$=$10^{11}$ to $N$=$10^{15}$ and the conclusion is almost the same. Note that if the full parameter optimization is also implemented to Refs.~\cite{ma2012statistical,yu2013generalized}, all the key rates will be almost the same. These results, once again, highlight the importance of full parameter optimization in the practical implementation of decoy-state MDI-QKD.

\subsection{Key rate comparison between full optimization and partial optimization}
\begin{table*}[hbt]
\centering
\begin{tabular}{c | @{\hspace{0.1cm}} c @{\hspace{0.1cm}} c @{\hspace{0.1cm}} c | @{\hspace{0.1cm}} c @{\hspace{0.1cm}} c  @{\hspace{0.1cm}} c| c @{\hspace{0.1cm}} c  @{\hspace{0.1cm}} c}
  \hline
   Distance  & 0km & 0km & 0km & 50km & 50km & 50km & 100km & 100km & 100km\\
   \hline
  Data-size  & $10^{12}$ & $10^{14}$ & $10^{18}$ & $10^{12}$ & $10^{14}$ & $10^{18}$ & $10^{12}$ & $10^{14}$ & $10^{18}$ \\
  \hline
  Unbiased   & $1.50\times10^{-5}$ & $3.98\times10^{-5}$ & $6.37\times10^{-5}$ & $3.21\times10^{-7}$ & $2.39\times10^{-6}$ &$ 5.71\times10^{-6}$ & 0              &$ 9.88\times10^{-8}$ & $4.72\times10^{-7}$\\
  \hline
  Simplified & $2.05\times10^{-5}$ & $6.27\times10^{-5}$ & $2.03\times10^{-4}$ & $3.36\times10^{-7}$ & $3.97\times10^{-6}$ & $1.66\times10^{-5}$ & 0                & $1.28\times10^{-7}$ & $1.21\times10^{-6}$  \\
  \hline
  Optimal    & $6.83\times10^{-5}$ & $1.72\times10^{-4}$ & $2.72\times10^{-4}$ & $1.68\times10^{-6}$ & $1.05\times10^{-5}$ & $2.24\times10^{-5}$ & $6.05\times10^{-10}$ & $4.61\times10^{-7}$ & $1.78\times10^{-6}$ \\
  \hline
\end{tabular}
\caption{Key rate values with different basis choices. The key rates are simulated with two decoy states and numerical approach. Unbiased denotes the standard protocol with equal basis choice; Simplified denotes the simplified choice with the (biased) basis choice independent of intensity choice; Optimal denotes the optimal choice with the (biased) basis choice depending on intensity choice. In a large data-set of $10^{18}$ (approaching asymptotic case), the key rates with optimal choice are around 300\% higher than those of unbiased choice and close to those of simplified choice. In a reasonable data-set ($10^{12}$ to $10^{14}$), the key rates with optimal choice are around 300\% higher than those of unbiased choice and around 200\% higher than those of simplified choice. This shows that the optimal choice can significantly increase the key rates in a practical setting with finite data-set.} \label{Tab:key:differentchoices}
\end{table*}

Table~\ref{Tab:key:differentchoices} shows the comparison results for different choices of bases. The key rates are simulated using the numerical method with two decoy states. Unbiased denotes the standard protocol with equal basis choice; Simplified denotes the simplified choice with basis choice independent of intensity choice; Optimal denotes the optimal choice with basis choice depending on intensity choice. In a larger data-set of $10^{18}$ (approaching asymptotic case), the key rates with optimal choice are around 300\% higher than those of unbiased choice and close to those of simplified choice. In a reasonable data-set ($N$=$10^{12}$ to $10^{14}$), the key rates with optimal choice are around 300\% higher than those of unbiased choice and around 200\% higher than those of simplified choice. Therefore, the optimal choice of parameters can significantly increase the key rates in a practical setting with finite data-set.

\section{Simulations on different number of decoy states} \label{Sec:numerical2}
\begin{figure}[!t]
\centering
\resizebox{8cm}{!}{\includegraphics{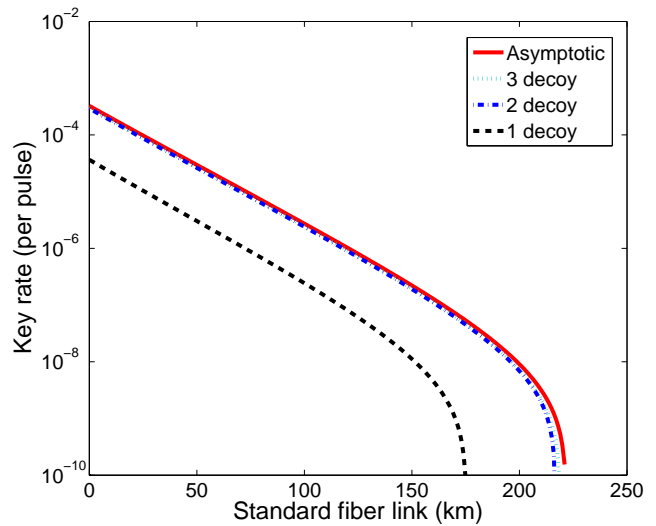}} \caption{(Color online) Asymptotic key rates with different number of decoy states. The solid curve is the one with infinite decoy states. The dashed, dashed-dotted, dotted curves are respectively the one, two, three decoy-state results using numerical methods (see Appendix~\ref{Sec:numerical}). The signal state $\mu$ is optimized in all cases, while some reasonable values of decoy states are adopted: for one decoy state, $\nu$=0.0005; for two decoy states, $\nu$=0.01 and $\omega$=0.0005; for three decoy states, $\nu_1$=0.1, $\nu_2$=0.01 and $\omega$=0.0005. We emphasize that the key rates with analytical methods of Appendix~\ref{Sec:analytical} are almost overlapped with the ones presented in this figure, which shows that the analytical approaches provide a highly good estimation. The estimation using two decoy states gives a nearly similar key rate to the one with three decoy states and is higher than one decoy-state case. Therefore, two decoy states can already result in a near optimal estimation and more decoy states cannot improve the key rate.} \label{Fig:key1}
\end{figure}

\begin{figure}[!t]
\centering
\resizebox{8cm}{!}{\includegraphics{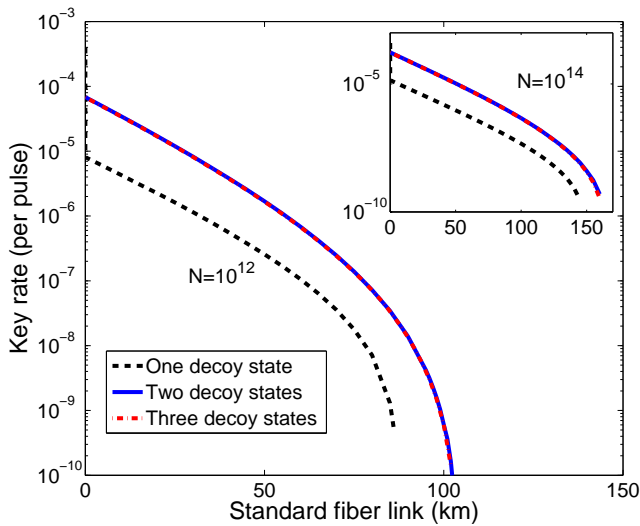}} \caption{(Color online) Secret key rate in logarithmic scale as a function of the distance under different numbe of decoy states. The main figure is for data-set $N$=$10^{12}$ and the inserted figure is for $N$=$10^{14}$. The key rates are obtained using numerical methods with one (dashed curve), two (solid curve), three (dashed-dotted curve) decoy states. The key rates with two and three decoy states are almost overlapped. In simulation, we perform a full parameter optimization for all cases. Our results show that after a full parameter optimization, the two decoy-state method can give an almost optimal key rate, which is higher than the one with one decoy state. Three decoy states cannot help to increase the key rate.} \label{Fig:key2:numerical12}
\end{figure}

\begin{table*}[hbt]
\centering
\begin{tabular}{c | @{\hspace{0.1cm}} c @{\hspace{0.1cm}} c @{\hspace{0.1cm}} c | @{\hspace{0.1cm}} c @{\hspace{0.1cm}} c  @{\hspace{0.1cm}} c| c @{\hspace{0.1cm}} c  @{\hspace{0.1cm}} c}
  \hline
   Distance & 0km & 0km & 0km & 50km & 50km & 50km & 100km & 100km & 100km\\
   \hline
  Data-size & $10^{12}$ & $10^{14}$ & $\infty$ & $10^{12}$ & $10^{14}$ & $\infty$ & $10^{12}$ & $10^{14}$ & $\infty$ \\
  \hline
  1decoy-Num & $8.10\times10^{-6}$ & $1.41\times10^{-5}$ & $3.59\times10^{-5}$ & $2.56\times10^{-7}$ & $9.34\times10^{-7}$ &$ 3.03\times10^{-6}$ & 0 &$ 4.64\times10^{-8}$ & $2.41\times10^{-7}$\\
  \hline
  2decoy-Num & $6.83\times10^{-5}$ & $1.72\times10^{-4}$ & $3.25\times10^{-4}$ & $1.68\times10^{-6}$ & $1.05\times10^{-5}$ & $2.95\times10^{-5}$ & $6.05\times10^{-10}$ & $4.61\times10^{-7}$ & $2.98\times10^{-6}$  \\
  2decoy-Ana & $6.65\times10^{-5}$ & $1.67\times10^{-4}$ & $3.25\times10{-4}$ & $1.67\times10^{-6}$ & $1.01\times10^{-5}$ & $2.95\times10^{-5}$ & $5.97\times10^{-10}$ & $4.48\times10^{-7}$ & $2.77\times10^{-6}$ \\
  \hline
  3decoy-Num & $6.76\times10^{-5}$ & $1.71\times10^{-4}$ & $3.03\times10^{-4}$ &  $1.66\times10^{-6}$ & $1.04\times10^{-5}$ & $2.92\times10^{-5}$ & $6.02\times10^{-10}$ & $4.55\times10^{-7}$ & $2.70\times10^{-6}$ \\
  \hline
\end{tabular}
\caption{Optimal key values under different number of decoy states. With finite data-set, the key rates with two decoy states are around one order of magnitude higher than the ones with one decoy state. Three decoy states cannot help to improve the key rates. Hence, two decoy states can achieve a near optimal key rate. In the case of two decoy states, the numerical method (Num) can only improve the key around 2\% over the ones using our analytical method (Ana). This shows that the two decoy-state analytical method presented in Appendix~\ref{Sec:analytical} can also result in a near optimal estimation.} \label{Tab:key:results}
\end{table*}

The simulation results using numerical methods (Appendix~\ref{Sec:numerical}) for different number of decoy states are shown in Figs.~\ref{Fig:key1},~\ref{Fig:key2:numerical12} and Table~\ref{Tab:key:results}. In the asymptotic case (Fig.~\ref{Fig:key1}), the key rate with two decoy states is close to the one with three decoy states as well as infinite decoy states and it is also larger than that with one decoy state. In a practical setting with finite data-set (Figs.~\ref{Fig:key2:numerical12}), the statistical fluctuations are simulated using the standard error analysis method~\cite{ma2005practical}. A full parameter optimization is conducted using our LSA. Some selected values of key rates are shown in Table~\ref{Tab:key:results}. Our results show that after a full parameter optimization, two decoy states can give an almost optimal key rate, which is much higher than the one with one decoy state. Three decoy states cannot improve the key rates much. Notice that the key rates using the analytical methods of Appendix~\ref{Sec:analytical} are almost overlapped with the ones using numerical methods (see Table~\ref{Tab:key:results} for the case of two decoy states). This shows that the analytical approaches provide a good decoy-state estimation. We have also performed simulations using the rigorous finite-key analysis presented in~\cite{marcos:finite:2013} and find that all the conclusions are the same. Therefore, in practical MDI-QKD, two decoy states combined with full parameter optimization can achieve a near optimal decoy-state estimation.

\section{Conclusion}~\label{Sec:conclusion}
In summary, we have shown the importance of full parameter optimization in practical decoy-state MDI-QKD and presented a novel LSA to realize such optimization. Full parameter optimization can increase the key rate around 200\% over the simplified choice~\cite{zhiyuan:experiment:2013}. LSA can be four orders of magnitude faster than a trivial exhaustive search to achieve a similar optimal key rate. In practice, implementing full parameter optimization requires slightly complex modulation schemes, as the sender's two modulators on the intensities and the bit information are dependent. However, one can in principle jointly modulate the two modulators using a single quantum random number generator~\cite{xu2012ultrafast}, which is similar to the setup in~\cite{Tittel:2012:MDI:exp}. Future research can also explore this full parameter optimization in an asymmetric setting of MDI-QKD~\cite{Feihu:practical}. Moreover, we have found that two decoy states already enable a near optimal decoy-state estimation in both asymptotic and finite-data case. Experimentalist can readily implement MDI-QKD by using our two decoy-state (analytical or numerical) approach combined with a full parameter optimization to enjoy the optimal system performance.

\section{Acknowledgments}
We thank C.~Lim, X.~Ma, J.~Ng, B.~Qi, especially M.~Curty, S.~Sun, Z. Tang and Z. Zhang, for enlightening discussions. We also thank P.~Roztocki for the preliminary work on the linear programming. Support from funding agencies NSERC, the CRC program, Connaught Innovation fund is gratefully acknowledged. F. Xu thanks Shahid U.H. Qureshi Memorial Scholarship for the support.

\appendix

\section{Finite decoy-state methods}

\subsection{Numerical approaches} \label{Sec:numerical}
Ignoring statistical fluctuations temporally, the estimations on $Y_{11}^{{\rm Z},L}$ and $e_{11}^{{\rm X},U}$ from Eq.~(\ref{Eqn:general:decoy}) are constrained optimisation problems, which is linear and can be efficiently solved by linear programming (LP). The numerical routine to solve these problems can be written as:
\begin{equation} \label{Eqn:decoy:numerical} \nonumber
\begin{aligned}
min: \space & Y_{11}^{\rm Z}, \\
s.t.: \space & 0\leq Y_{nm}^{\rm Z}\leq 1, with \ n,m\in {\mathcal S}_{\rm cut} \\
& Q_{q_aq_b}^{\rm Z}-(1-\sum_{n,m\in S_{cut}}e^{-(q_a+q_b)}\frac{q^n_a}{n!}\frac{q^m_b}{m!})\leq \\
& \sum_{n,m\in {\mathcal S}_{\rm cut}}e^{-(q_a+q_b)}\frac{q^n_a}{n!}\frac{q^m_b}{m!}Y_{nm}^{\rm Z}\leq Q_{q_aq_b}^{\rm Z} \\ \\
Max: \space &e_{11}^{\rm X}, \\
s.t.: \space & 0\leq Y_{nm}^{\rm X}\leq 1, 0\leq Y_{nm}^{\rm X}e_{nm}^{\rm X}\leq 1, with \ n,m\in {\mathcal S}_{\rm cut} \\
       & Q_{q_aq_b}^{\rm X}-(1-\sum_{n,m\in {\mathcal S}_{\rm cut}}e^{-(q_a+q_b)}\frac{q^n_a}{n!}\frac{q^m_b}{m!})\leq \\
       & \sum_{n,m\in S_{cut}}e^{-(q_a+q_b)}\frac{q^n_a}{n!}\frac{q^m_b}{m!}Y_{nm}^{\rm X}\leq Q_{q_aq_b}^{\rm X} \\
       & Q_{q_aq_b}^{\rm X}E_{q_aq_b}^{\rm X}-(1-\sum_{n,m\in {\mathcal S}_{\rm cut}}e^{-(q_a+q_b)}\frac{q^n_a}{n!}\frac{q^m_b}{m!})\leq \\
       & \sum_{n,m\in S_{cut}}e^{-(q_a+q_b)}\frac{q^n_a}{n!}\frac{q^m_b}{m!}Y_{nm}^{\rm X}e_{nm}^{\rm X}\leq Q_{q_aq_b}^{X}E_{q_aq_b}^{X}
\end{aligned}
\end{equation}
where ${\mathcal S}_{\rm cut}$ denotes a finite set of indexes $n$ and $m$, with ${\mathcal S}_{\rm cut}=\{n,m \in {\mathbb N} \ {\rm with}\ n\leq{}N_{\rm cut} and m\leq{}M_{\rm cut} \}$,
for prefixed values of $N_{\rm cut}\geq{}2$ and $M_{\rm cut}\geq{}2$. In our simulations, we choose $N_{\rm cut}=7$ and $M_{\rm cut}=7$, as larger $N_{\rm cut}$ and $M_{\rm cut}$ have negligible effect on decoy-state estimation. More discussions can be seen in~\cite{ma2012statistical}. Here, $q\in\{\mu,\nu\}$ for one decoy-state estimation; $q\in\{\mu,\nu,\omega\}$ for two decoy-state estimation; $q\in\{\mu,\nu_1,\nu_2,\omega\}$ for three decoy-state estimation. Notice that statistical fluctuations can be easily conducted by adding constraints on the experimental measurements of $Q_{q_aq_b}^{\lambda}$ and $E_{q_aq_b}^{\lambda}$. These additional constraints can be analyzed by using statistical estimation methods, such as standard error analysis~\cite{ma2012statistical} or Chernoff bound~\cite{marcos:finite:2013}. A rigorous finite-key analysis can also be implemented by following the technique presented in~\cite{marcos:finite:2013}.

\subsection{Analytical approaches} \label{Sec:analytical}
\subsubsection{One decoy state}
We consider an estimation method with only one decoy state $\nu$ satisfying $\mu>\nu$. Our starting point is Eq.~(\ref{Eqn:general:decoy}). To estimate $Y^{{\rm Z},L}_{11}$, we use \emph{gaussian elimination}. Firstly, we simultaneously cancel out all the third order terms $Y_{12}$, $Y_{21}$, $Y_{30}$, $Y_{03}$:
\begin{multline} \label{1decoy:Eq1}
\mu^3\times Q_{\nu\nu}^{\rm Z}e^{2\nu}-\nu^3\times Q_{\mu\mu}^{\rm Z}e^{2\mu}= \\
 \mu^2\nu^2(\mu-\nu)Y_{11}^{\rm Z}+\mu^3(Y_{00}^{\rm Z}+\nu Y_{01}^{\rm Z}+\nu Y_{10}^{\rm Z}+\nu^2Y_{02}^{\rm Z}/2+\nu^2Y_{20}^{\rm Z}/2)\\
-\nu^3(Y_{00}^{\rm Z}+\mu Y_{01}^{\rm Z}+\mu Y_{10}^{\rm Z}+\mu^2Y_{02}^{\rm Z}/2+\mu^2Y_{20}^{\rm Z}/2) + \\
\sum_{n+m>3}^{\infty}\frac{(\nu^{n+m}\mu^3-\mu^{n+m}\nu^3)}{n!m!}Y_{nm}^{Z} \leq \\
\mu^2\nu^2(\mu-\nu)Y_{11}^{\rm Z}+\mu^3(Y_{00}^{\rm Z}+\nu Y_{01}^{\rm Z}+\nu Y_{10}^{\rm Z}+\nu^2Y_{02}^{\rm Z}/2+\nu^2Y_{20}^{\rm Z}/2)
\end{multline}
where the inequality comes from the fact that $(\nu^{n+m}\mu^3-\mu^{n+m}\nu^3)<0$ for $n+m>3$. Next, from $Q_{\nu\nu}^{Z}E_{\nu\nu}^{\rm Z}$, we have
\begin{multline} \label{1decoy:Eq2}
Q_{\nu\nu}^{\rm Z}E_{\nu\nu}^{\rm Z} e^{2\nu}= \sum_{n,m=0}^{\infty}\frac{\nu^{n+m}}{n!m!}Y_{nm}^{\rm Z}e_{nm}^{\rm Z}\geq \\
Y_{00}^{Z}e_{00}^{\rm Z}+\nu Y_{01}^{Z}e_{01}^{\rm Z}+\nu Y_{10}^{Z}e_{10}^{\rm Z}+\nu^2Y_{02}^{\rm Z}e_{02}^{Z}/2+\nu^2Y_{20}^{\rm Z}e_{20}^{\rm Z}/2 \\
= (Y_{00}^{\rm Z}+\nu Y_{01}^{\rm Z}+\nu Y_{10}^{\rm Z}+\nu^2Y_{02}^{\rm Z}/2+\nu^2Y_{20}^{\rm Z}/2)/2
\end{multline}
where the final equality is from $e_{0m}^{\rm Z}$=$e_{n0}^{\rm Z}$=1/2, which is a standard assumption in QKD descending from the fact that the error rate cause by 0-photon pulse is 1/2. Therefore, by combining Eq.~(\ref{1decoy:Eq1}) and Eq.~(\ref{1decoy:Eq2}), we have a lower bound for $Y_{11}^{\rm Z}$
\begin{eqnarray} \label{Eq:1decoy:Y11}
Y_{11}^{\rm Z} \geq Y_{11}^{{\rm Z},L}=\frac{\mu^3 Q_{\nu\nu}^{\rm Z}e^{2\nu}(1-2E_{\nu\nu}^{\rm Z})-\nu^3Q_{\mu\mu}^{\rm Z}e^{2\mu}}{\mu^2\nu^2(\mu-\nu)}
\end{eqnarray}

To estimate $e^{{\rm X},U}_{11}$, we use the same method as \cite{Feihu:practical,marcos:finite:2013} and obtain an upper bound for $e^{\rm X}_{11}$
\begin{multline} \label{Eq:1decoy:e11}
e^{\rm X}_{11} \leq e_{11}^{{\rm X},U}=\frac{1}{(\mu-\nu)^2Y^{{\rm X},L}_{11}}\times \\
(e^{2\mu}Q^{\rm X}_{\mu\mu}E^{\rm X}_{\mu\mu} + e^{2\nu}Q^{\rm X}_{\nu\nu}E^{\rm X}_{\nu\nu} - e^{\mu+\nu}Q^{\rm X}_{\mu\nu}E^{\rm X}_{\mu\nu}-e^{\nu+\mu}Q^{\rm X}_{\nu\mu}E^{\rm X}_{\nu\mu})
\end{multline}

\subsubsection{Two decoy states}
We consider an estimation method with two decoy states $\nu$, $\omega$ satisfying $\mu>\nu>\omega\geq0$. We have the lower bound $Y_{11}^{Z,L}$ and the upper bound $e^{{\rm X},U}_{11}$~\cite{Feihu:practical,marcos:finite:2013}
\begin{multline} \label{Eq:2decoy:Y11}
Y_{11}^{{\rm Z},L}=\frac{1}{(\mu-\omega)^2(\nu-\omega)^2(\mu-\nu)}\times \\
[(\mu^2-\omega^2)(\mu-\omega)(Q_{\nu\nu}^{\rm Z}e^{2\nu}+Q_{\omega\omega}^{\rm Z}e^{2\omega}-Q_{\nu\omega}^{\rm Z}e^{\nu+\omega}-Q_{\omega\nu}^{\rm Z}e^{\omega+\nu}) \\ -(\nu^2-\omega^2)(\nu-\omega)(Q_{\mu\mu}^{\rm Z}e^{2\mu}+ Q_{\omega\omega}^{\rm Z}e^{2\omega}-Q_{\mu\omega}^{\rm Z}e^{\mu+\omega}-Q_{\omega\mu}^{\rm Z}e^{\omega+\mu})],
\end{multline}
\begin{multline} \label{Eq:2decoy:e11}
e^{{\rm X},U}_{11}=\frac{1}{(\nu-\omega)^2Y^{{\rm X},L}_{11}}\times \\
[e^{2\nu}Q_{\nu\nu}^{\rm X}E_{\nu\nu}^{\rm X}+e^{2\omega}Q_{\omega\omega}^{\rm X}E_{\omega\omega}^{\rm X}-e^{\nu+\omega}Q_{\nu\omega}^{\rm X}E_{\nu\omega}^{\rm X}-e^{\omega+\nu}Q_{\omega\nu}^{\rm X}E_{\omega\nu}^{\rm X}].
\end{multline}

\section{Experimental implementation of the one decoy-state method}~\label{App:1decoy}
\begin{table*}[hbt]
\centering
\begin{tabular}{c @{\hspace{0.5cm}} c @{\hspace{0.5cm}} c @{\hspace{0.5cm}} c @{\hspace{0.5cm}} c @{\hspace{0.5cm}} c  @{\hspace{0.5cm}} c}
\hline
Systematic parameters & $\eta_{d}$    & $e_{d}$    & $Y_{0}$               & $L$    & $\epsilon$          & $N$ \\
           & 8.2\%         & 0.8\%        & $5\times 10^{-5}$  & 10km   & $2.7\times10^{-3}$  & $1.11\times10^{11}$ \\
\hline
One decoy state  & $P_{s}$ & $Q^{\rm Z}_{\mu\mu}$    & $E^{\rm Z}_{\mu\mu}$      & $Y^{{\rm Z},L}_{11}$         & $e^{{\rm X},U}_{11}$       & $R_{L}$ \\
                 & 0.45     &$5.33\times10^{-6}$  & 4.03\%                & $4.17\times10^{-5}$    & 12.97\%   & \textbf{$4.26\times10^{-10}$} \\
\hline
\end{tabular}
\caption{Experimental parameters and results. These experimental parameters include the detection efficiency $\eta_{d}$, the total misalignment error $e_{d}$, the background rate $Y_{0}$, the transmission distance from Alice to Bob $L$, the security bound $\epsilon$ and the total number of signals sent by Alice and Bob. $P_{s}$ is the optimal probability to send a signal state. $R_{L}$ denotes the lower bound of the key rate. } \label{Tab:exp:results}
\end{table*}

Here we experimentally implement the one decoy-state method (Appendix~\ref{Sec:analytical}) in a polarization-encoding MDI-QKD system presented in~\cite{zhiyuan:experiment:2013}. Using the systematic parameters shown in Table~\ref{Tab:exp:results}, we perform a numerical optimization to maximize the key rate. The optimal intensities of the signal and the decoy state are respectively around $\mu=0.1$ and $\nu=0.01$ and the optimal probability to send a signal state is around $P_{s}=0.45$. Here we choose the simplified choice with the same probability to select $\rm Z$ or $\rm X$ basis, which is simpler for our implementation. An optimal choice will be implemented in our future experiments.

We test the one decoy-state method over 10km standard telecom fiber and operate the system at a repetition rate around 500 kHz. The system is operated for 55 hours and a total number of signals around $N=1.11\times 10^{11}$ is generated. The experimental results are shown in Table~\ref{Tab:exp:results}. Around 50 secure keys are generated. Our results demonstrate the possibility of one decoy-state method. This method is highly simple to implement in a practical MDI-QKD system, but gives a slightly lower key rate.

One might ask `why the key rate is low in our implementation?'. This is mainly due to the low repetition rate of our system as well as the finite-key effect. The key generation rate can be substantially improved by increasing the repetition rate: First, more pulses can be sent out in a reasonable time frame, leading to tighter bounds in the decoy-state estimation. Second, by using four detectors rather than two (as our implementation), we can get at least a four-fold increase in the key rate. Third, given a larger data size, we can reduce the portion of pulses sent as decoy states and more pulses can be sent out in signal states for key generation. The speed of our system is limited by the performance of our SPDs. Our simulation (see the black dashed curves for the asymptotic case in Fig.~\ref{Fig:key1} and the finite-key case with $N=10^{14}$ in Fig.~\ref{Fig:key2:numerical12}) shows that, with commercial four single-photon detectors having say an efficiency of about 15\%~\footnote{For instance, id210/220, manufactured by IDQ, can have over 20\% efficiency and a gate rate over 100 MHz.}, the key generation rate can be around $10^{-5}$ per pulse at say 30km fiber. If gating up to 100 MHz, the key rate (with the finite key effect) can be up to 1 kbps. Moreover, by using state-of-the-art super-conducting single photon detectors with over 90\% quantum efficiency, the key rate can be as high as 100 kbps. This high-speed system will be implemented in our future experiment.

\section{Local search algorithm}~\label{App:Local}
\begin{figure}
\centering \resizebox{6cm}{!}{\includegraphics{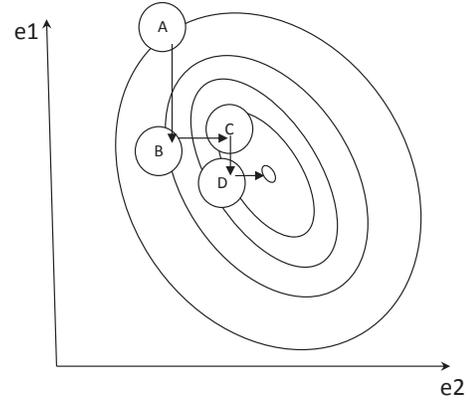}}
\caption{Coordinate Descent (CD). CD algorithm searches along one coordinate direction in each iteration, and it use a different coordinate directions cyclically. For instance, on the equiv-error
contour of two dimensional subspace, CD starts at point A (arbitrarily) and descents vertically along the direction $e_1$ to B, then horizontally along the direction $e_2$ toward C.
After cyclic iterations of vertical and horizontal descent, the algorithm stops at D where it is very close to the optimal. This simplified two-dimensional example illustrates how generalized search in any
dimensional space can be done analogously.
}\label{Fig:coordinateDescent}
\end{figure}

In the decoy-state approach to BB84 and MDI-QKD, the key rate depends largely on the systematic parameters and the optimal key rate is achieved via numerical optimization on many dimensions (parameters). To reduce both computational time and storage space of this optimization process, local search algorithm (LSA), a combination of coordinate descent (CD) and backtrack search (BS) algorithm, is adopted in lieu of the conventional exhaustive search algorithm.

There are salient drawbacks of exhaustive search. If the search is too fine,
the computational time and space are challenging. If the search is too coarse, we will miss finer details. In contrast, CD is a non-derivative approximation to the well-known steepest descent (SD) algorithm~\cite{luo1992convergence}. This approximation is necessary from the facts that our key rate is an implicit function (a linear
program) of the parameters and the actual finding of gradients and hessians of SD cannot be done easily. CD converges to the same optimal point as SD, even though it requires more iterations. CD can fix
low-speed by making large progress at the start, and it can also fix
in-accuracy by re-defining how close to the optimal point the algorithm
can stop. Table~\ref{Tab:localsearch} compares the
speed and accuracy of exhaustive search versus LSA.

CD is based on the idea that the minimization of a multivariate
function (key rate $R$) can be achieved by minimizing it along one
direction at a time (see Fig.~\ref{Fig:coordinateDescent}). Instead of varying descent direction according to gradient, one fixes descent directions at the
outset~\cite{bertsekas1999nonlinear}. These directions are usually the
cartesian bases, \ie, $e_i$ with $i$=1,2,3,.... In the two
decoy-state case, $e_1$=$\mu$, ..., $e_4$=$P_{\mu}$, ..., $e_7$=$P_{X|\mu}$,
$e_8$=$P_{X|\nu}$, and this basis is iterated through one at a time,
minimizing the objective function with respect to the current
coordinate direction. Mathematically, to optimize $\mu$, if $\mu^{k}$ (optimized $\mu$ in the $k$th iteration) is given, the
minimization of key rate $R$ (see Eq.~\ref{Eqn:Key:formula}) along $\mu$ coordinate in the $k+1$th
iteration is:
\begin{multline} \label{Eq:LS}
\mu^{k+1}=arg \max_{y \in R}
R(P_{\mu}^{k+1},P_{\nu}^{k+1}, \\
P_{X|\mu}^{k+1}, P_{X|\nu}^{k+1}, P_{X|\omega}^{k+1}, y, \nu^{k}, \omega^{k})
\end{multline}

By doing line search in each iteration, we automatically have a
sequence of vector $x_0,x_1,x_2,...$, where
$x_i=((P_{\mu})_i,(P_{\nu})_i,...,(P_{X|\mu})_i,...,(\nu)_i,(\omega)_i)$
and the sequence of key rate: $R(x_0) \geq
R(x_1) \geq R(x_2) \geq ...$.

The demonstration of convexity in Appendix~\ref{App:convex} will make the result of LSA a global optimum. Although CD requires an intelligent guess to start with, the starting point in convex topologies can be in theory any non-zero objective (key rate) point in the search
space. In practice, prior research can shed light on the choices of
initial parameters, and these parameters often are good candidates for the starting guess.

After a direction along a coordinate chosen in CD, we still
have to do a one-dimensional line search problem to compute how far the search
can move along a given coordinate. This is realized via the BS algorithm. BS starts at the end of previous iterations,
and makes progress toward a minima along the chosen coordinate
direction. With a step from one side to the minima to the other
side, the algorithm found a turning point. From there, BS searches
backward again toward the minima until the same turning point is
found with greater accuracy. The procedure is iterated until
converged to the minima. At this point in the search space, the CD
algorithm restarts with a new direction of line-search.

\section{Other practical aspects}

\subsection{Optimal signal state in the asymptotic case}~\label{App:optimalmu}
\begin{figure}[hbt]
\centering \resizebox{6cm}{!}{\includegraphics{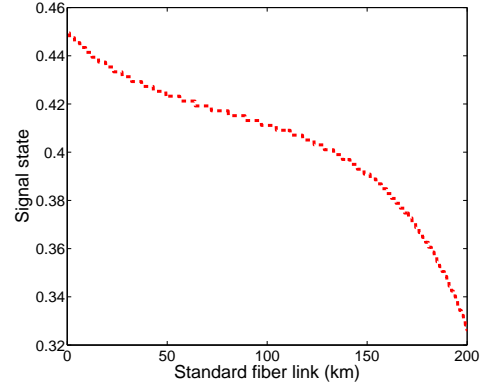}}
\caption{Plot of the intensity $\mu$ as a function of the transmission distance for the decoy-state MDI-QKD with infinite decoy states.}
\label{AppendixA:fig:DecoyGen}
\end{figure}

We consider the asymptotic case with infinite decoy states. From the model presented in~\cite{Feihu:practical}, we have:
\begin{equation}\label{Eqn:Asym}
\begin{aligned}
e_{11}^{\rm X} &\approx e_d \\
Y_{11}^{\rm Z} &\approx t_at_b\eta_d^2 \\
Q^{\rm Z}_{\mu\mu} &\approx \mu^2t_at_b\eta_d^2(1+2e_d)/2 \\
E^{\rm Z}_{\mu\mu} &\approx \frac{2e_d}{1+e_d}
\end{aligned}
\end{equation}
where $t_a$ ($t_b$) is the channel transmittance from Alice (Bob) to Charlie.

Substituting Eq.~(\ref{Eqn:Asym}) into Eq.~(\ref{Eqn:Key:formula}), the key generation rate is given by:
\begin{equation} \label{Eqn:Key:formula:simplify}
\begin{aligned}
R\geq \mu^2e^{-2\mu}t_at_b\eta_d^2[1-H_{2}(e_d)]- \\
\frac{\mu^2t_at_b\eta_d^2(1+2e_d)}{2}f_{e}(\frac{2e_d}{1+e_d})H_{2}(\frac{2e_d}{1+e_d}),
\end{aligned}
\end{equation}
The expression is optimized if we choose $\mu=\mu^{Optimal}$ which fulfills $\frac{\partial R}{\partial \mu}=0$:
\begin{equation} \label{Optmu:Decoymu}
(1-\mu)\exp(-2\mu)=\frac{f_e(\frac{2e_d}{1+e_d})(1+2ed)H_2(\frac{2e_d}{1+e_d})}{1-H_2(e_{d})}.
\end{equation}
By using $f_e(\cdot)=1.16$ and $e_d=1.5\%$, we solve Eq.~\eqref{Optmu:Decoymu} and get $\mu^{Optimal}=0.42$. The numerical simulation for the optimal $\mu$ at different distances is shown in Fig.~\ref{AppendixA:fig:DecoyGen}. We can see that Eq.~\eqref{Optmu:Decoymu} is a good approximation. Moreover, from Eq.~(\ref{Optmu:Decoymu}) and Fig.~\ref{AppendixA:fig:DecoyGen}, the optimal intensity is a continuous function with transmission distance and there is \emph{only} one solution for $\mu\in(0,1]$. That is, the key rate is a \emph{convex function} to $\mu$ at a fixed distance.

\subsection{Convex function of the key rate}~\label{App:convex}
\begin{figure*}[!t]
\centering \subfigure[] {\includegraphics [width=6cm]
{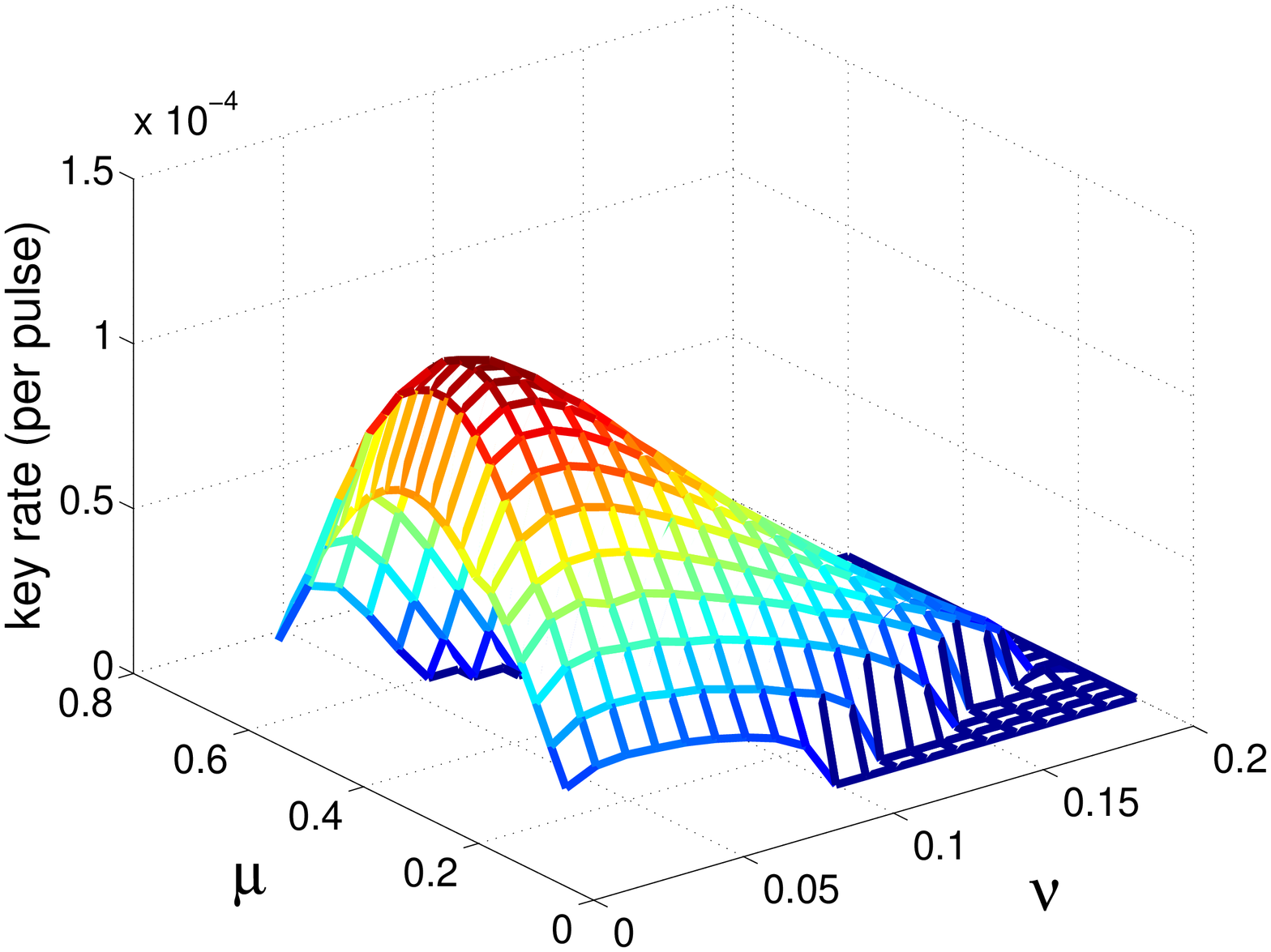} \label{Fig:convexity}} \qquad \subfigure[]
{\includegraphics [width=5cm] {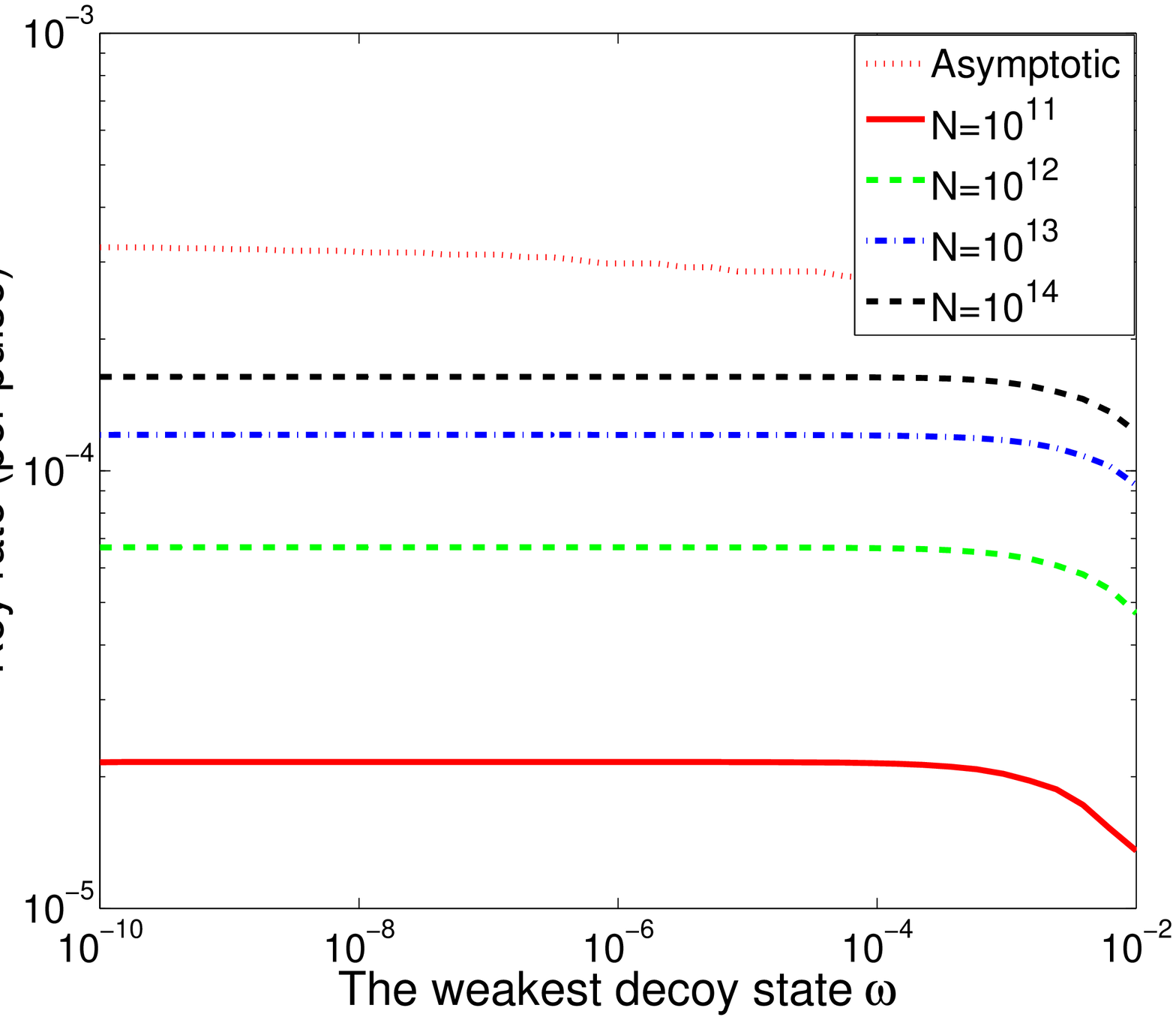}
\label{Fig:w}} \caption{ (Color online) (a) Convexity of key rate function. The key rate is simulated by sweeping $\mu$ and $\nu$, and optimizing other parameters. (b) Key rate as a function of the decoy state $\omega$. As long as the intensity of $\omega$ is below $1\times10^{-3}$, the key rate is very close to the optimum. A perfect vaccum ($\omega=0$) is not essentially required in practical decoy-state QKD experiments.}
\end{figure*}

The CD and BS algorithm~\cite{boyd2004convex} works in any general topology of the search space, but the saving in efficiency comes only when the underlying topology is a \emph{convex optimization} problem. It is our goal here to demonstrate the convexity of the key rate as a function of the parameters. Appendix~\ref{App:optimalmu} has already shown that the key rate is indeed convex in the case of infinite decoy states. In a practical setting with finite decoy states, for illustration purpose, we have chosen to sweep the intensities of $\mu$ and $\nu$ and optimized the other dimensions at 0km with $N$=$10^{12}$. Fig.~\ref{Fig:convexity} shows that the convexity of key rate function, which allows a unique optimal set of parameters to be employed in an actual experiment.

\subsection{The effect of the smallest decoy state $\omega$} \label{App:omega}
In practice, it is usually difficult to create a perfect vacuum state in decoy-state QKD experiments~\cite{rosenberg2007long,dixon2008gigahertz}. The different intensities are usually generated with an intensity modulator, which has a finite extinction ratio below 30 dB. Thus, the question is: what is the effect of the intensity of the smallest decoy states $\omega$ on the secure key rate? Here, we perform a simulation on the sweep of the smallest intensity $\omega$ in the case of two decoy states. The result is shown in Fig.~\ref{Fig:w}. We find that after a full optimization of parameters, the optimal $\omega$ is in the vicinity of $5\times10^{-6}$. As long as the intensity of $\omega$ is below $1\times10^{-3}$, the key rate is very close to the optimum. In summary, a perfect vaccum ($\omega=0$) is \emph{not} essentially required in practical decoy-state QKD experiments.

\bibliographystyle{apsrev4-1}
\bibliography{decoyMDI}

\end{document}